\documentclass[11pt,a4paper]{article}

\usepackage[T1]{fontenc}
\usepackage[utf8]{inputenc} 
\usepackage{graphicx}
\usepackage{amsmath,amssymb, bm}
\usepackage{cite}
\usepackage[colorlinks=true,linkcolor=blue,citecolor=blue,urlcolor=blue]{hyperref}
\usepackage{orcidlink}
\usepackage{QCDEX2026}

\newcommand{\rt}{{\bm r}}

\newcommand{\asb}{\bar{\alpha}_s}

\begin{document}

\noindent\includegraphics[width=0.4\textwidth]{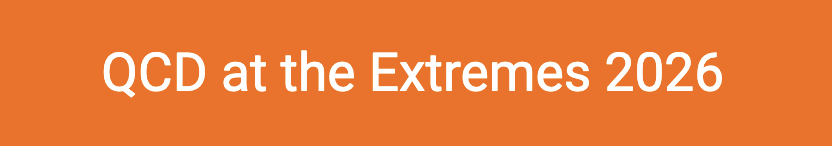}\par
\vspace{0.6em}
\begin{center}
  {\LARGE\bfseries Entanglement entropy in low x Deep Inelastic Scattering \par}
  \vspace{0.8em}
  {\large
    Martin Hentschinski\orcidlink{0000-0003-2922-7308}
\par}
  \vspace{0.5em}
  {\small
    Departamento de Actuaría, Física y Matemáticas, Universidad de las Américas Puebla, Ex-Hacienda Santa Catarina Martír S/N, 
San Andrés Cholula, 72820 Puebla, Mexico\par
}
  \vspace{0.35em}
  {\small $^{*}$Corresponding author: \href{mailto:martin.hentschinski@udlap.mx}{martin.hentschinski@udlap.mx}\par}
  \vspace{0.35em}
  {\small\itshape refereed by the conveners of QCDEX2026\par}
\end{center}
\vspace{0.5em}

\begin{abstract}
In a 2017 paper by Kharzeev and Levin, it has been proposed that hadronic entropy, extracted from the multiplicity of final state hadrons, provides a measure of entanglement in the proton wave function, observed in Deep Inelastic Scattering (DIS) of electrons and protons. In this contribution we discuss how the density matrix of the low x DIS system can give rise to a mixed state and therefore to finite entanglement entropy. We further present recent numerical results on the determination of low x dipole multiplicities, based on the solution of the Levin-Lublinsky evolution equation.
\end{abstract}

\noindent\textbf{Keywords:} QCD; quantum information; low $x$ evolution equations

\section{Introduction}

Deep Inelastic Scattering (DIS) of electrons on protons is a reaction which allows to probe the partonic content of the proton in experiment. Two variables which characterize the reaction are the photon virtuality $-q^2 = Q^2$ and Bjorken $x = Q^2/(2 p\cdot q)$, where $p$ is the light-like momentum of the incoming proton and $q$ the momentum of the exchanged virtual photon. Both variables allow to express the invariant mass of the produced hadronic system through $W^2 = (p+q)^2 =  \frac{1-x}{x}Q^2$. At high $Q^2$, the reaction is well described within the parton model, which treats the proton as a collection of quasi-free i.e. incoherent partons, which are subject to a classical probability distribution, the parton distribution functions or short PDFs. At the same time, both the scattering electron and the proton are pure quantum states, which are evolved with the unitary scattering operator. It is therefore a natural question to ask how both statements make sense at the same time and what this relation can teach us about the DIS reaction. 

This issue has been discussed in \cite{Kharzeev:2017qzs}, where the authors arrive at the following picture, see also the discussion in \cite{Hentschinski:2025luo,Hentschinski:2023nhy,Hentschinski:2023izh,Hentschinski:2022rsa,Hentschinski:2021aux}:  

\begin{itemize}
\item The proton is  a highly entangled quantum state; due to color confinement,  color charges are entangled to a degree such that the observation of a  single quark or gluon is impossible
\item DIS at large photon virtuality corresponds to a partial observation of this entangled state. 
\item Quantum mechanics teaches us that a partial observation of an entangled state gives rise to a mixed state with finite von Neumann entropy
\item A measure of the degree of entanglement is provided by entanglement entropy. The authors of \cite{Kharzeev:2017qzs} further argue that this entanglement entropy can be observed through hadronic entropy, $S_h = -\Sigma_N P_N \ln P_N$ where $P_N$ are hadron multiplicities. 

\end{itemize}
These ideas have been explored during the last years
\cite{Hentschinski:2025luo,Hentschinski:2023nhy,Hentschinski:2023izh,Hentschinski:2022rsa,Hentschinski:2021aux}:
Making use of $1+0$ dimensional dipole model, with free parameters
fixed from a comparison to PDFs, a good description of hadronic
entropy extracted by the H1 collaboration \cite{H1:2020zpd} has been
achieved.  The results are impressive since they allow to describe
features of the hadronic final state, i.e. its entropy, using only
information provided by the initial state, i.e. parton distribution
functions. This observation provides strong support for the idea that
one indeed observes  entanglement entropy, which is generated at
early times of the reaction, i.e. during the interaction of the proton
with the virtual photon. Open questions which still need to be
addressed are:
\begin{enumerate}
\item What are the entangled Hilbert spaces ${\cal H} = {\cal H}_A \otimes {\cal H}_B $, which give rise to a mixed state $\rho_A = {\rm tr}_B \rho_{AB}$ and finite entanglement entropy $S = -{\rm tr}\rho_A\ln \rho_A$?
\item The description within the $1+0$ dimensional dipole model requires adjustment of parameters for each bin in $Q^2$ and a natural question is whether entanglement entropy be described for all bins at once.
\item Last but not least, the relation to particle production in a Monte Carlo event  generator, the role of soft gluons as well as  hadronization. While we will not comment on this question in this contribution, we would like to draw the attention  to the recent article \cite{Hentschinski:2025pyq}, which starts to investigate those questions.
\end{enumerate}
It is needless to say that the entire discussion takes place in the context of the  emerging field of  quantum information science applied to high energy scattering, see  \cite{Bloss:2026yrf,Hatta:2026dqs,Braun:2026tga,Adelaja:2026qhq,Zheng:2026pma,Bondarenko:2026jkx,Fucilla:2026oal,Han:2026woa} for recent works in this field.

\begin{figure}[htbp]
  \centering
   \includegraphics[width=.4\textwidth]{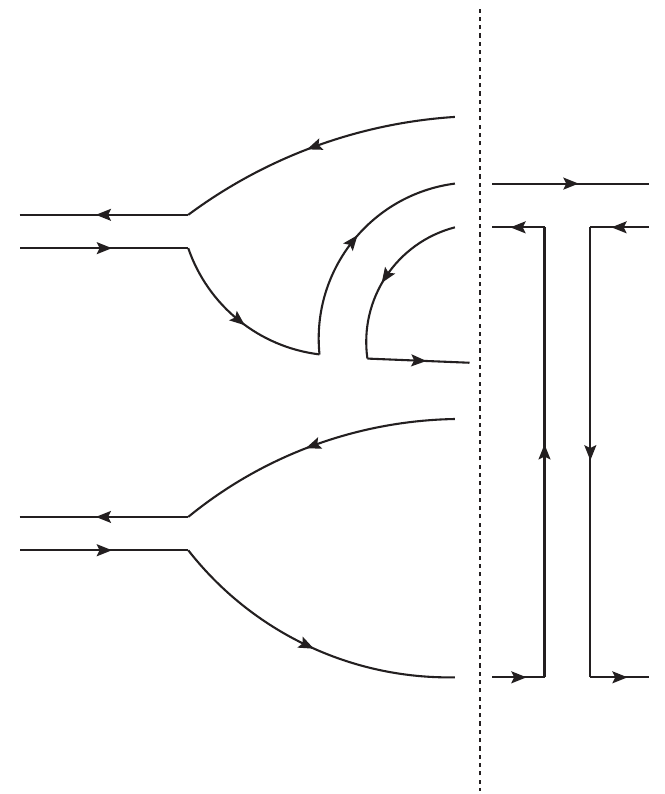}
 
  \caption{Scattering of two color dipoles, e.g.\ a virtual photon and a heavy onium state; the dashed line indicates the transition from the initial-state wave function towards the color-entangled state. Each of the scattering particles maintains its transverse position during the scattering, while its color charge is  modified.}
  \label{fig:schematic}
\end{figure}

\section{Formalism and methods}
To arrive at an improved understanding of these questions, we consider
the DIS process within a simplified model, the so-called dipole model \cite{Mueller:1993rr}, which
gives a good approximation to the reaction in the limit  $x \to 0$.
This limit is characterized by a high invariant mass of the hadronic system
and large hadronic mean multiplicities. Within this dipole model, the
virtual photon and the hadron are modeled through a system of color
dipoles, which are obtained through a rearrangement of a quark antiquark and $n$ gluon degrees of freedom into color dipoles, using a multi-color approximation. Each dipole is characterized by its
transverse separation as well as by its rapidity. Within a leading
logarithmic approximation, hadron and photon wave functions can  be expressed in a
dipole Fock basis, see also Fig.~\ref{fig:schematic} and
\cite{Hentschinski:2026otq} for details. Within high energy
factorization, the transverse positions of the color dipoles are conserved
during the interaction, while the interaction Hamiltonian is non-diagonal in color; one therefore arrives after the interaction at  a pure, but entangled
quantum state of the photon-hadron system. 

A possibility to arrive at a
reduced state, which is diagonal in dipole multiplicities, is a simple
Gaussian average over the color degrees of freedom of the color
dipoles. Indeed, due to color confinement, an observation of color is
not possible in experiment and as a consequence the relevant density
matrix is the reduced density matrix, with color degrees of freedom
traced over \cite{Hentschinski:2026otq}. While a realistic average over color degrees of freedom involves the complete hadronization process, a Gaussian average appears to be an appropriate treatment within a high energy leading logarithmic approximation. If such an average is implemented, entanglement
entropy is then obtain directly in terms of the dipole multiplicities
$p_n$ through $S = - \sum_n p_n \ln p_n$, where we refer the interested reader again to  \cite{Hentschinski:2026otq} for the complete details.

To determine probabilities for the  observation of $n$ color dipoles, it is possible to formulate within a leading logarithmic approximation evolution equations, which determine their dependence on  $Y=\ln 1/x$. Such an evolution equation has been formulated by Levin and  Lublinsky   \cite{Levin:2003nc} for probability densities which are unintegrated with respect to the transverse size of dipoles, but with the longitudinal momentum dependence integrated out. These probability densities satisfy the normalization condition,
\begin{align}
    \label{eq:probs_dipole}
\sum_{n=1}^\infty  \prod_{i=1}^n \int d^2\rt_i  P_n(Y; \rt_1, \rt_2, \ldots, \rt_n) & =1,
\end{align}
and their evolution equation reads
\begin{align}
  \frac{dP_n(Y; \rt_1, \ldots, \rt_n)}{dY}
  = & - \sum_{i=1}^n \sigma(\rt_i)\, P_n(Y;\rt_1, \ldots, \rt_n) \nonumber\\
  & + \sum_{i=1}^{n-1} K(\rt_i, \rt_n)\, P_{n-1}(Y; \rt_1, \ldots, \rt_i + \rt_n, \ldots, \rt_{n-1}).
  \label{eq:dpn}
\end{align}
Here
\begin{align}
  \label{eq:kernel}
  K(\rt_i, \rt_n) &= \frac{\asb}{2\pi} 
  \frac{(\rt_i + \rt_n)^2}{\rt_i^2 \, \rt_n^2} \,
  \theta(\rt_i^2 - \rho^2)\, \theta(\rt_n^2 - \rho^2),
  &
  \asb & = \frac{\alpha_s N_c}{\pi},
\end{align}
is the dipole splitting  kernel supplemented by an ultraviolet (UV) regulator $\rho$;  it denotes the probability for a dipole of size  $(\rt_{i}+\rt_{n})$ to decay into two dipoles of sizes $\rt_{i}$ and $\rt_{n}$. 
Within a more complete description, the dipole probabilities densities still need to be convoluted with the wave function of external scattering particles. These wave functions provide  finite support only in a limited region of dipole sizes, roughly $1/\Lambda_{\rm QCD} > |\rt_i| > 1/Q$. Since our current study does not include those wave functions, we take these effects into account through setting the UV cut-off to $\rho = 1/Q$ in the following and through using a  infrared cut-off, which we identify with the size of the initial dipole.

The virtual correction or the dipole survival probability can be finally obtained as an integral over the real emission kernel:
\begin{align}
  \label{eq:integraged_kernel}
  \sigma(\rt_i) & = \int d^2 \rt'K(\rt_i-\rt', \rt') 
  \approx \overline{\alpha}_s\ln \frac{\rt_i^2}{\rho^2}.
\end{align}
 As the initial conditions we assume that the proton consists of a single dipole of size $\rt$ at $Y = 0$
\begin{align}
    P_n(Y=0; \rt_1, \ldots, \rt_n) & = \delta_{n1} \delta^{(2)}(\rt_1 -\rt).
\end{align}
 The mean dipole multiplicity is finally defined as
\begin{align}
    \label{eq:mean_dipole}
\langle n\rangle(x,r^2/\rho) &=  \sum_{n=1}^\infty  n  \prod_{i=1}^n \int d^2\rt_i  P_n(Y; \rt_1, \rt_2, \ldots, \rt_n),
\end{align}
where the dependence of the $P_n$ on the initial conditions and cut-offs is understood. 

\section{Results}

\begin{figure}[!ht]
    \centering     
    \includegraphics[width=1.0\linewidth]{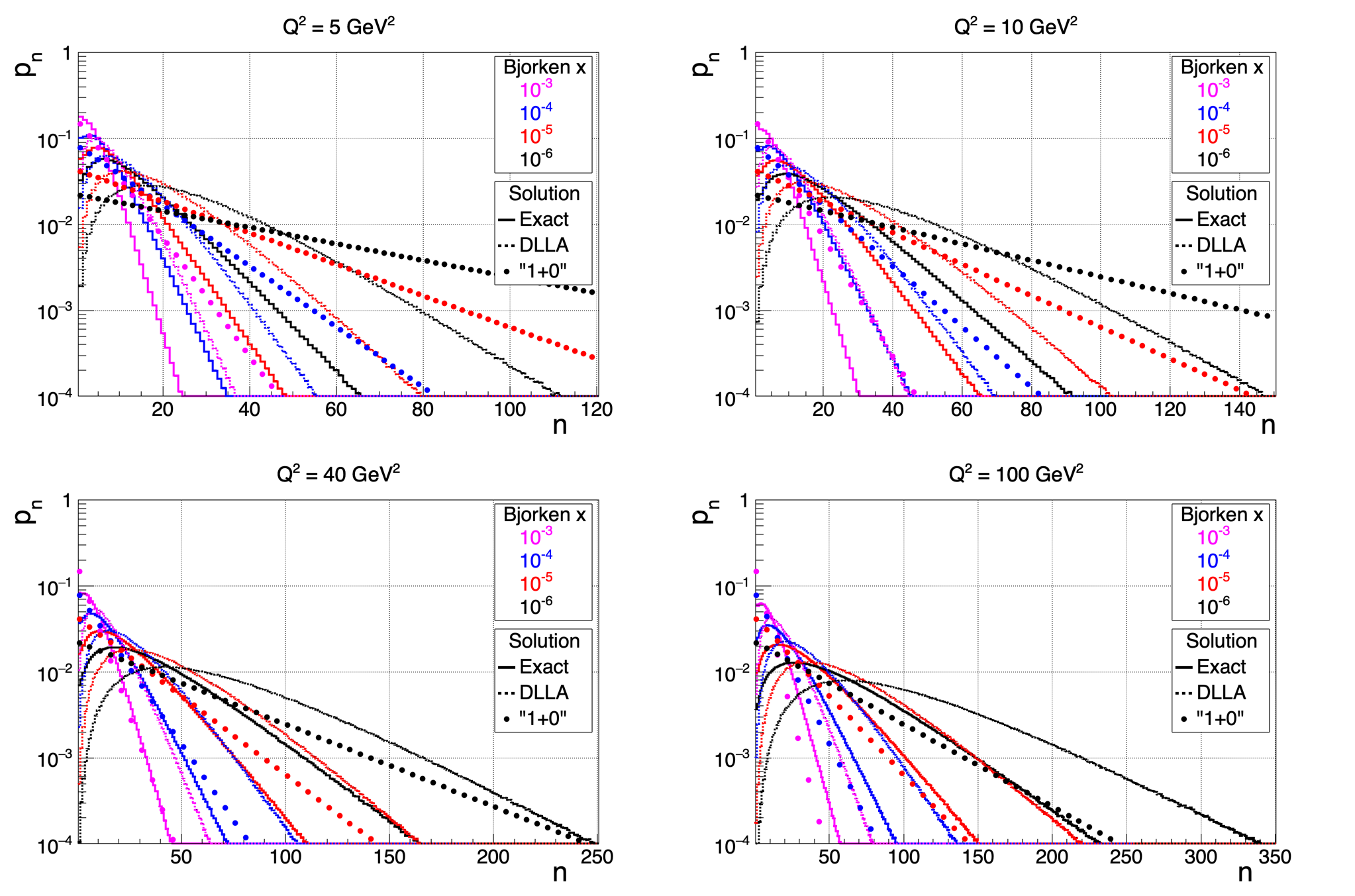}
    \caption{Monte Carlo results from {\sf DIPMAR} for probability distributions $p_n$ of a number of generated dipoles $n$
     corresponding to the fixed $\asb = 0.1$ and the initial dipole size (for `Exact' and `DLLA') 
     $r=0.85\,$fm.
     In the `Exact' case, the IR cut-off $\Delta_{\rm IR} = r = 0.85\,$fm has been used.
     }
    \label{fig:Pn}
\end{figure}
To numerical determine the dipole multiplicities, two Monte Carlo codes  named \texttt{DIPMAR} and \texttt{LLMC} have been developed, see \cite{Hentschinski:2026otq} for full details. Fig.~\ref{fig:Pn} show the integrated dipole multiplicities and we compare also to an implementation of the simplified 1+0 dimensional model, which has been used in previous studies. We find that the maxima of the distribution moves for the Levin Lublinsky evolution equation towards finite $n$, which is an important feature also found in H1 data, see e.g. \cite{H1:2020zpd}.
\begin{figure}[!ht]
    \centering     
    \includegraphics[width=1.0\linewidth]{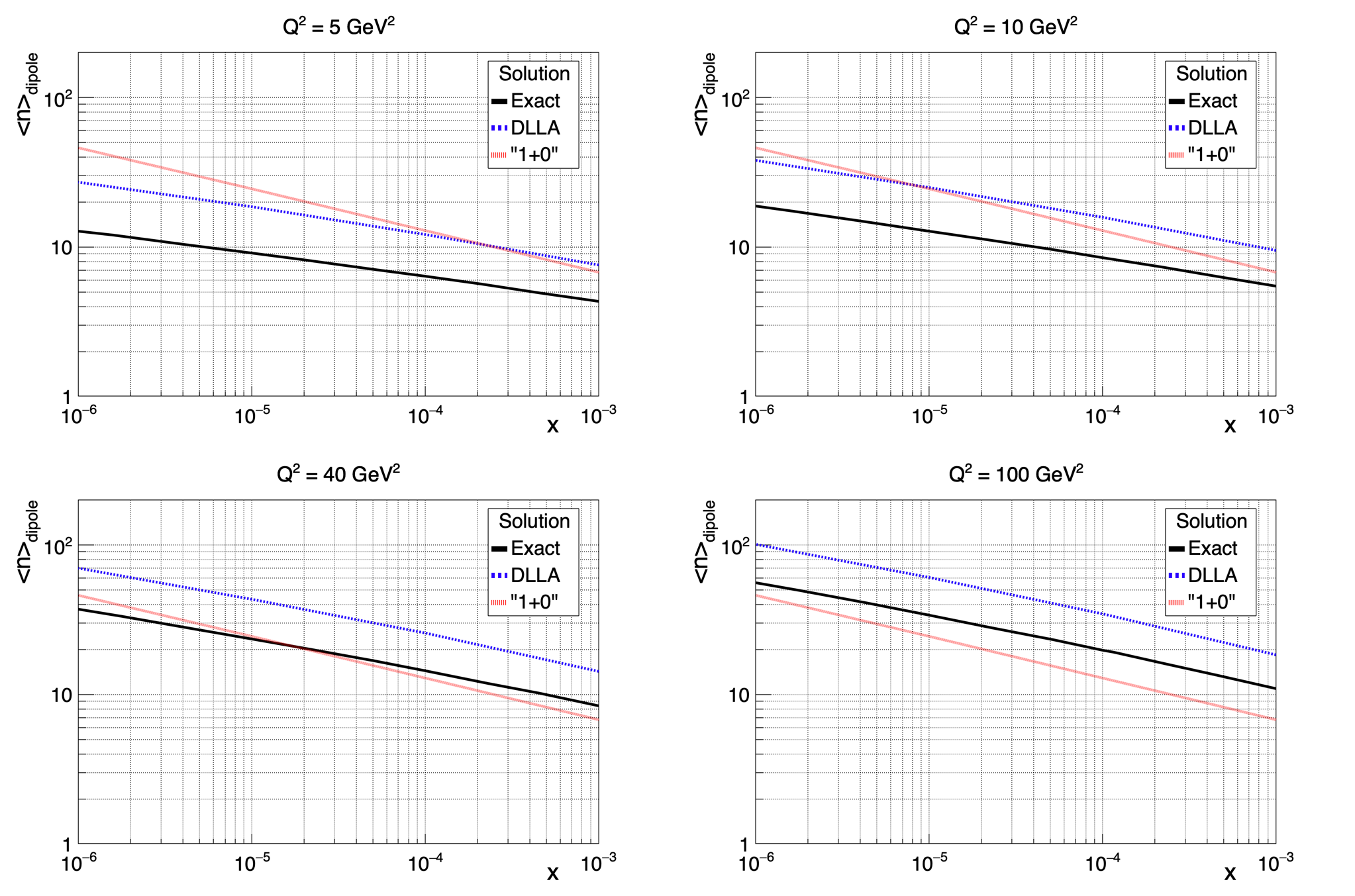}
    \caption{Monte Carlo results from {\sf DIPMAR} for the dipole mean multiplicity $\langle n \rangle_{\rm dipole}$ 
     corresponding to the fixed $\asb = 0.1$ and the initial dipole size (for `Exact' and `DLLA') $r=0.85\,$fm.
     In the `Exact' case, the IR cut-off $\Delta_{\rm IR} = r = 0.85\,$fm was also applied.
    }
    \label{fig:nmeandipole}
\end{figure}
Fig.~\ref{fig:nmeandipole} presents the mean multiplicity. For a double logarithmic version of the evolution equation, which again is presented in full detail in \cite{Hentschinski:2026otq}, we find a very precise agreement with the analytic fixed coupling result $\langle n\rangle = I_0 \left(2\sqrt{ \frac{\alpha_s N_c}{\pi} \ln \frac{r^2}{ \rho^2} Y} \right)$. To have realistic values for the mean dipole number (in the range of phenomenological gluon distributions), we use for this study the value $\alpha_s N_c/{\pi} = 0.1$, which is small, but which can be justified through treatments of the QCD running coupling constant in NLO BFKL fits , see e.g. \cite{Hentschinski:2012kr,Hentschinski:2013id,ArroyoGarcia:2019cfl,Hentschinski:2025ovo,Hentschinski:2026nfe}. 
\begin{figure}[!ht]
    \centering     
    \includegraphics[width=1.0\linewidth]{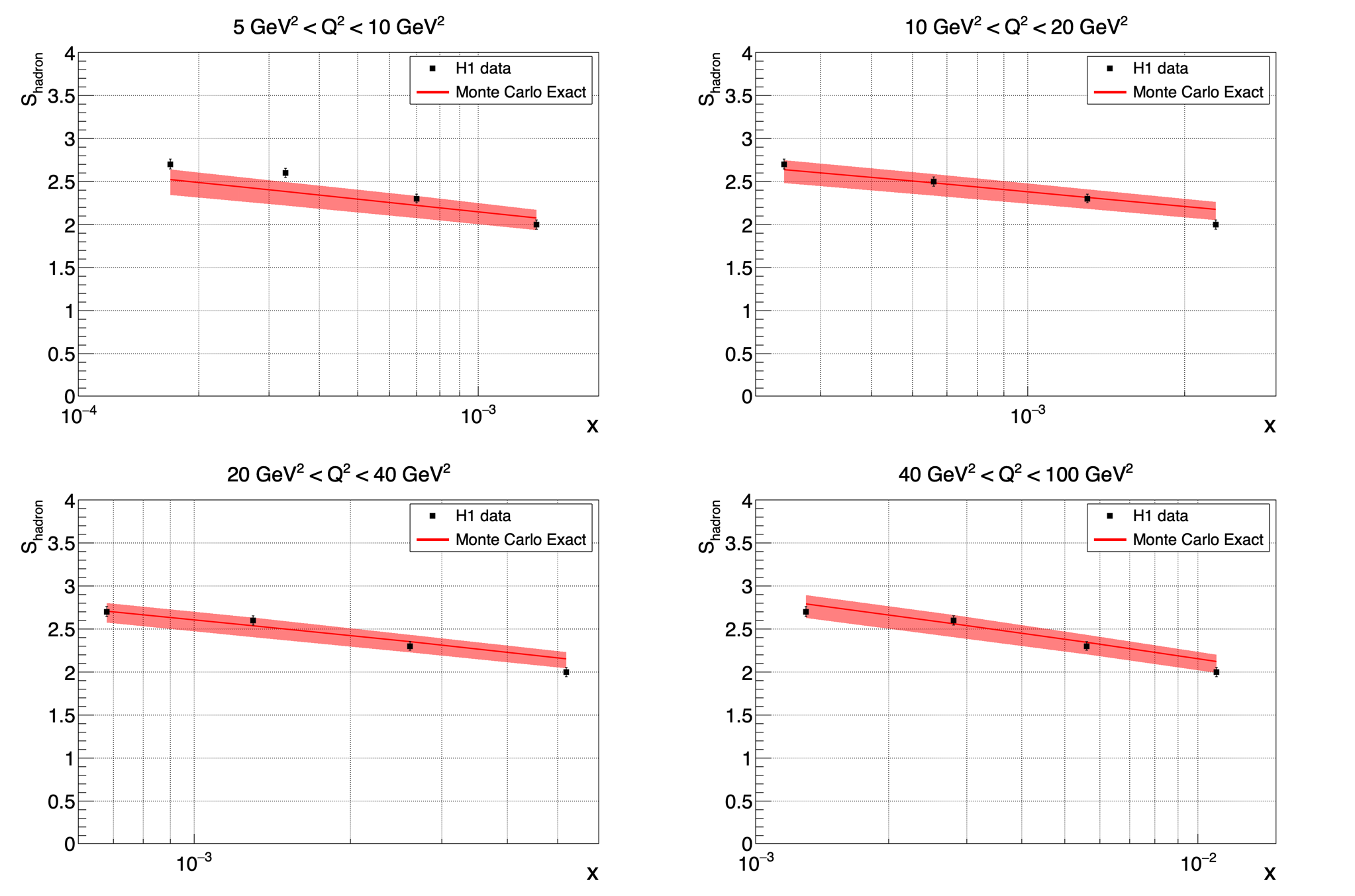}
    \caption{Comparison of Monte Carlo results from {\sf DIPMAR} with the H1 data for the hadron entropy $S_{\rm hadron}$. 
     The numerical results have been obtained through the MCMC solution of the exact dipole evolution equation with 
     the fixed $\asb = 0.1$ and the initial dipole size $r=1.0\,$fm (with $\Delta_{\rm IR} = r$). 
     The uncertainty bands correspond to the $Q^2$ ranges of the H1 measurements.}
   \label{fig:Sdipole}
\end{figure}
In Fig.~\ref{fig:Sdipole} we finally present our results for the dipole entanglement entropy, including a comparison to H1 data \cite{H1:2020zpd}. For this comparison we need to take into account that  in experiment one only measure the charged hadron multiplicity, while our implementation provides probabilities for all hadrons. Assuming that the majority of produced hadrons are pions and that pions $\pi^\pm, \pi^0$ are produced with equal probability, we determine the charged hadron probability $P_N$ from the dipole multiplicities $p_n$ through
\begin{align}
    P_N & = \sum_{n=N}^\infty\binom{n}{N} q^N (1-q)^{n-N} p_n, \qquad q = \frac{2}{3},
    \label{eq:PN}
\end{align}
see also the discussion in \cite{Hentschinski:2022rsa}.  We finally explore to which extend our implementation is in agreement with the proposal of maximal entanglement, see \cite{Kharzeev:2017qzs}: for a Hilbert space of dimension $d$, the density operator of a maximally mixed state is provided by $\hat{\rho} = \hat{1\!\!1}/d$ and gives rise to von Neumann entropy in the Boltzmann form $S = -{\rm tr} \hat{\rho} \ln \hat{\rho} = \ln d$. For an infinite dimensional Hilbert space, maximal entanglement is only realized in a sub-space with a dimension that is of the order of two times the mean multiplicity, $d \simeq 2 \langle n\rangle$. Within a 1+0 dimensional model one finds \cite{Hentschinski:2022rsa,Hentschinski:2023izh} indeed that dipole multiplicity is  for large dipole multiplicities $S_{\rm 1+0} \simeq \ln  \langle n\rangle + 1 = \ln  \langle e n\rangle $, with $e \simeq 2.718$ the Euler number, which is numerically close to a factor of two; for a comparison to the charged hadron multiplicity this then turns into  $S_{\rm 1+0} \simeq  \ln e \tfrac{2}{3}\langle  n\rangle $. For the present  solution of the Levin-Lublinsky, a comparison to the  exact results yields  $S_{\rm LL} \simeq  \ln \tfrac{2}{3} \langle  n\rangle + 0.85$. With $e^{0.85}\simeq 2.33965$, we find a slightly different constant, which reflects differences in the resulting Hilbert space, while the central interpretation remains valid: the effective dimension of the Hilbert space is roughly related through a factor of two to the mean multiplicity; color and momentum/coordinate Hilbert spaces are maximally entangled.  

\section{Summary and outlook}
In this talk we presented results on the study of entanglement entropy
in the context of Deep Inelastic Scattering of electrons and protons
in the low $x$ limit. In the produced partonic system, color and 
momentum degrees of freedom are entangled. Since color confinement
does not allow for the observation of colored degrees of freedom, a
natural candidate for the the reduced density matrix is provided through an average over the color Hilbert space. We provide a numerical solution to the Levin-Lublinsky equation, through two Monte Carlos \texttt{DIPMAR} and \texttt{LLMC},  based on a fixed and small running coupling
constant. As previously our study suggests that we are dealing with an almost maximally entangled system. 
 Future studies should address in more detail the role of
soft gluons, see \cite{Hentschinski:2025pyq} for first steps in this
direction.

\section*{Acknowledgements}
I am very grateful to Krysztof Kutak, Wies{\l}aw P{\l}aczek, and Martin Rohrmoser for collaboration; in particular for their invaluable work in the  development of the  the Monte Carlo codes  \texttt{DIPMAR}  and  \texttt{LLMC} which enabled the presented numerical results. Support  through the  SECIHTI project CBF-2026-49 is gratefully acknowledged. 

\bibliographystyle{QCDEX2026}
\bibliography{QCDEX2026_Hentschinski}

\end{document}